\documentclass[aps,prl,twocolumn,superscriptaddress,english]{revtex4}
\usepackage{amssymb}
\usepackage{graphicx}
\usepackage{amsmath}
\usepackage{amsthm}
\usepackage{amsfonts}
\usepackage[T1]{fontenc}
\usepackage[latin9]{inputenc}
\usepackage{array}
\usepackage{multirow}
\usepackage{color}
\usepackage{esint}
\usepackage{bm}
\usepackage{color}
\usepackage{bbm}
\usepackage{hyperref}
\usepackage{babel}
\usepackage{titlesec}
\usepackage{hyperref}
\hypersetup{breaklinks,
colorlinks=true,%
citecolor=blue,%
filecolor=black,%
linkcolor=blue,%
urlcolor=blue}
\makeatletter

\def\Arg{\mathop{\operator@font Arg}\nolimits}

\begin{document}
%\linenumbers

\global\long\def\id{\mathbbm{1}}
\global\long\def\ui{\mathbbm{i}}
\global\long\def\ud{\mathrm{d}}

\title{A universal pairing gap measurement proposal by dynamical excitations in 2D doped attractive Fermi-Hubbard model with spin-orbit coupling}

\author{Huaisong Zhao$^{\#}$}
\affiliation{College of Physics, Qingdao University, Qingdao 266071, China}
\author{Rui Han$^{\#}$}\thanks {$^{\#}$These authors contributed equally to this work.}
\affiliation{College of Physics, Qingdao University, Qingdao 266071, China}
\author{Ling Qin}
\affiliation{College of Physics and Engineering, Chengdu Normal University, Chengdu 611130, China}
\author{Feng Yuan}\thanks{Corresponding author: yuan@qdu.edu.cn}
\author{Peng Zou}\thanks{phy.zoupeng@gmail.com}
\affiliation{College of Physics, Qingdao University, Qingdao 266071, China}

%\date{\today}% It is always \today, today,

%\settocdepth{part}
\begin{abstract}
By calculating dynamical structure factor of two-dimensional doped attractive Fermi-Hubbard model with Rashba spin-orbit coupling, we not only investigate collective modes and single-particle excitations of the system during the phase transition between Bardeen-Cooper-Schrieffer superfluid and topological superfluid, but also propose a universal method to measure pairing gap measurement in an optical lattice system. Our numerical results show that the area of the molecular excitation peak at the transferred momentum ${\bf q}=\left[\pi,\pi\right]$ is proportional to the square of the pairing gap in the system with Rashba SOC. In particular, this method is very sensitive to the pairing gap. This goes on verifying that this method is universal to measure the pairing gap in a doped optical lattice with Rashba SOC. These theoretical results are important for experimentally measuring the pairing gap and studying the topological superfluid in an optical lattice.

\end{abstract}

\maketitle

{\it Introduction.}--Since the experimental realization of Raman-type spin-orbit coupling (SOC) effect in bosonic and fermionic quantum gases \cite{Liu2009,Lin2011s,Cheuk12,Wang12,Burdick16,Huang16,Liang2023}, the SOC effect becomes a controllable and tunable strategy, and has been widely used to study some interesting quantum phases and new quasiparticles, such as topological superfluid \cite{Yang12,Liuxj2014}, topological Fulde-Ferrell-Larkin-Ovchinnikov phase \cite{Cao14,Zhou14} and Majorana fermions\cite{xu14,Liu12}, Majorana solitons \cite{Zou16}, rashbons \cite{Jayantha11}, Weyl fermions \cite{Gong11,Wang2021,Lu2020}. Due to the heating effect of laser light, the superfluid state of SOC Fermi gases is still a challenging issue in experiments. Many researchers put forward some theoretical proposals of Rashba SOC \cite{Dalibard11,Xu13,Anderson13,Zhou19,Zhai15}. Theoretically the phase transition between a conventional Bardeen-Cooper-Schrieffer (BCS) superfluid and a topological superfluid has been studies in SOC Fermi superfluid by continuously varying the strength of a magnetic field. Something most interestingly is that the topologically non-trivial superfluid state is accompanied by the appearance of long-sought Majorana fermions, which are brought by the topological boundary or impurity and are believed to be the essential quantum bits for topological quantum computation. So the experimental preparation and confirmation of topological superfluid is an important scientific research issue. Compared with BCS superfluid, the pairing gap in topological superfluid is small, and the band structure becomes complex owing to effect of the Zeeman field or SOC effect. Therefore, it is not easy to check the superfluid by measuring the pairing gap, and  distinguish topological superfluid from other types of superfluid.

Dynamical excitations can play an important role in understanding properties of both BCS and topological superfluids. Physically all dynamical excitations information can be well displayed by the dynamical structure factor, which is a many-body physical quantity and can be experimentally measured directly by the two-photon Bragg spectroscopy in ultracold quantum gases or inelastic neutron-scattering experiments in condensed matter physics \cite{Hoinka17,Biss2022,Senaratne2022,Li2022,Hong2017,Ran2017}. Numerically the quantum Monte Carlo had been used to study dynamical structure factors \cite{Vitali2022}. Many theoretical papers had already considered the conventional superfluids and discussed their collective modes and single-particle excitations \cite{Combescot2006,Zou2018,Zou2021,Zhao2020,Kuhnle2010,Watabe2010}. Some strategies had been proposed and realized to measure pairing gap. For instance, the pairing gap can be obtained by the horizontal threshold energy of the single-particle excitations, $\omega=2\Delta$, which indicates the minimum energy to break a Cooper pair \cite{Hoinka17}. However, this method is fail to work in the SOC superfluid due to its complex structure of excitation spectrum. Moreover, only few papers discuss the collective excitations under SOC effect \cite{He2013,Zhao2023,Gao2023,Koinov2017}, and lack of systematic study from both collective excitations and single-particle excitations.

The repulsive and attractive Fermi-Hubbard model in an optical lattice are widely used to simulate the many-body physics in condensed matter physics \cite{Greif13,Parsons16,Cheuk16,Boll16,Arovas2022}. Experimentally the attractive Fermi-Hubbard model in ultracold atoms had been realized \cite{Mitra2018,Peter2020,Hackermuller2010,Gall2020,Hartke2023}. Moreover, the physical properties of this attractive Fermi-Hubbard model without SOC had also been studied theoretically \cite{Cocchi2016,Moreo07,Shenoy2008}. T. Hartke {\it et al.} experimentally observed the formation and spatial ordering of nonlocal fermion pairs in the pseudo-gap regime of an attractive system by using a two-species gas of degenerate $^{40}$K atoms \cite{Hartke2023}. So the pairing gap measurement in this system receives significant attention. In our previous work (arXiv:2305.09685), we studied the dynamical excitations of the attractive Fermi-Hubbard model based on random phase approximation (RPA) theory, and proposed a new method to measure the pairing gap by calculating the molecular excitations at the transfer momentum $q=[\pi,\pi]$. It is interesting to check the universality of this method in other complex system, for example, the SOC Fermi superfluid in an optical lattice. In this paper, we numerically calculate the dynamical structure factor within RPA theory, and discuss the influence of SOC effect to dynamical excitations, which includes the evolution of collective modes and single-particle excitations in phase transition between BCS superfluid and topological superfluid. Significantly, we will discuss a universal strategy to measure pairing gap in SOC Fermi-Hubbard mode by dynamical excitations.

{\it The model.}-- In this paper, we consider an attractive Fermi-Hubbard mode with Rashba SOC effect in a 2D square optical lattice, whose interaction can be successfully tuned by Feshbach resonances \cite{Frohlich2011}. Interestingly a topological superfluid can exist in this system with a proper Rashba SOC strength and an external Zeeman field.
 Therefore, this mode becomes a good platform to study the physical properties of the topological Fermi superfluid in lattice system.
The Hamiltonian of this model can be described as
\begin{widetext}
\begin{eqnarray}\label{Humodel2}
 H =&-&t\sum_{<jl>}\Psi_{j\sigma}^{\dagger}\Psi_{l\sigma}
 -\left(\mu+h\sigma_{z}\right)\sum_{j\sigma}\Psi_{j\sigma}^{\dagger}\Psi_{j\sigma}
 -U\sum_{j}\Psi_{j\uparrow}^{\dagger}\Psi_{j\downarrow}^{\dagger}\Psi_{j\downarrow}\Psi_{j\uparrow} \nonumber \\
 &+&\lambda\sum_{j}\left[\Psi_{j+x\downarrow}^{\dagger}\Psi_{j\uparrow}-\Psi_{j+x\uparrow}^{\dagger}\Psi_{j\downarrow}
+i\Psi_{j+y\downarrow}^{\dagger}\Psi_{j\uparrow}+i\Psi_{j+y\uparrow}^{\dagger}\Psi_{j\downarrow}+h. c.\right].
\end{eqnarray}
\end{widetext}
Here $\langle jl \rangle$ denotes all possible nearest-neighbour sites of the 2D lattice with hopping energy strength $t$. $\mu$ is the chemical potential, and $h$ is the perpendicular Zeeman field strength. $\sigma_{z}$ is the Pauli matrix. $\lambda$ is the strength of the Rashba SOC. $\Psi_{j\sigma}^{\dagger}(\Psi_{j\sigma})$ is the atom creation (annihilation) operator with spin $\sigma$ ($\sigma=\uparrow,\downarrow$) in lattice $j$. $U$ is the strength of on-site attraction interaction, which is used as the typical energy unit in our following discussion, together with lattice constant $a_{0}$ as the length unit. In this paper, we take a typical parameter $t/U=0.3$, and a temperature is almost close to zero ($T/U=0.001$).

When the system comes into the superfluid,
we define a pairing order parameter $\Delta=U\langle \Psi_{\downarrow}\Psi_{\uparrow} \rangle$. In the frame of mean-field theory,  the four-operator term in the interaction Hamiltonian can be dealt with $U\Psi^{\dagger}_{\uparrow}\Psi^{\dagger}_{\downarrow}\Psi_{\downarrow}\Psi_{\uparrow} \approx\Delta\Psi^{\dagger}_{\uparrow}\Psi^{\dagger}_{\downarrow}+\Delta^{*}\Psi_{\downarrow}\Psi_{\uparrow}$.
Then the Hamiltonian of Eq. \ref{Humodel2} has a simplified mean-field form in momentum representation
\begin{eqnarray}\label{tjbm}
H_{\rm mf}&=&\sum_{{\bf k},\sigma}\left(\xi_{\bf k}-h\sigma_{z}\right)\Psi^{\dagger}_{{\bf k}\sigma}\Psi_{{\bf k}\sigma}\nonumber\\
&-&\sum_{{\bf k}}\left(\Delta^{*}\Psi_{{\bf k}\downarrow}\Psi_{-{\bf k}\uparrow}+\Delta \Psi^{\dagger}_{-{\bf k}\uparrow}\Psi^{\dagger}_{{\bf k}\downarrow}\right)\nonumber\\
&+&\sum_{\bf k}\left[\lambda_{\rm so}({\bf k})\Psi^{\dagger}_{{\bf k}\uparrow}\Psi_{{\bf k}\downarrow}+h.c.\right],
\end{eqnarray}
where $\xi_{\bf k}=-Zt\gamma_{\bf k}-\mu$ is the single particle energy with $\gamma_{\bf k}=(\cos{k_{x}}+\cos{k_{y}})/2$,  and $Z=4$ is the nearest lattice number in this 2D square lattice system. Also the SOC term in momentum representation reads $\lambda_{\rm so}({\bf k})=\lambda(\sin{k_{x}}+i\sin{k_{y}})$.

The mean-field Hamiltonian in Eq.~\ref{tjbm} can be solved by motion equations of Green's functions \cite{Zhao2023-2}, together with several Green's functions, including the spin-up Green's function $G_{1}({\bf k},\tau-\tau^{'})=-\langle T \Psi_{{\bf k}\uparrow}(\tau)\Psi^{\dagger}_{{\bf k}\uparrow}(\tau^{'})\rangle$  and spin-down one $G_{2}({\bf k},\tau-\tau^{'})=-\langle T \Psi_{{\bf k}\downarrow}(\tau)\Psi^{\dagger}_{{\bf k}\downarrow}(\tau^{'})\rangle$.
Also the singlet pairing Green's function is  $\Gamma^{\dagger}({\bf k},\tau-\tau^{'})=-\langle T \Psi^{\dagger}_{{\bf -k}\downarrow}(\tau)\Psi^{\dagger}_{{\bf k}\uparrow}(\tau^{'})\rangle$.
Several extra Green's functions are brought by the SOC effect, and
 we have to define spin flip Green's function $S^{\dagger}({\bf k},\tau-\tau^{'})=-\langle T \Psi_{{\bf k}\downarrow}(\tau)\Psi^{\dagger}_{{\bf k}\uparrow}(\tau^{'})\rangle$, spin-up triplet pairing Green's function
$F_{1}^{\dagger}({\bf k},\tau-\tau^{'})=-\langle T \Psi^{\dagger}_{{\bf -k}\uparrow}(\tau)\Psi^{\dagger}_{{\bf k}\uparrow}(\tau^{'})\rangle$, and spin-down triplet pairing Green's function  $F_{2}^{\dagger}({\bf k},\tau-\tau^{'})=-\langle T \Psi^{\dagger}_{{\bf -k}\downarrow}(\tau)\Psi^{\dagger}_{{\bf k}\downarrow}(\tau^{'})\rangle$. It should be noted that $G_{1}\neq G_{2}$ and $F_{1}^{\dagger}\neq F_{2}^{\dagger}$ due to a non-zero Zeeman field $h$. The definition of these six Green's functions comes from their coupling effect during solving motion equations of Green's function. The expressions of these six Green's functions in the momentum representation read
\begin{eqnarray}\label{Greenf}
G_{1}({\bf k},\omega)=\sum_{a=1,2}\left[\frac{U'^2_{a{\bf k}}}{\omega-E_{a{\bf k}}}+
\frac{V'^2_{a{\bf k}}}{\omega+E_{a{\bf k}}}\right],\nonumber\\
G_{2}({\bf k},\omega)=\sum_{a=1,2}\left[\frac{U^2_{a{\bf k}}}{\omega-E_{a{\bf k}}}+
\frac{V^2_{a{\bf k}}}{\omega+E_{a{\bf k}}}\right],\nonumber\\
\Gamma^{\dagger}({\bf k},\omega)=\sum_{a=1,2}\left[
\frac{\alpha_{a{\bf k}}}{\omega+E_{a{\bf k}}}+\frac{\beta_{a{\bf k}}}{\omega-E_{a{\bf k}}}\right],\nonumber\\
S^{\dagger}({\bf k},\omega)=\sum_{a=1,2}\left[\frac{\lambda^{*}_{\rm so}({\bf k})P_{a{\bf k}}}{\omega-E_{a{\bf k}}}+
\frac{\lambda^{*}_{\rm so}({\bf k})Q_{a{\bf k}}}{\omega+E_{a{\bf k}}}\right],\nonumber\\
F_{1}^{\dagger}({\bf k},\omega)=\sum_{a=1,2}\left[\frac{\lambda^{*}_{\rm so}({\bf k})T'_{a{\bf k}}}{\omega-E_{a{\bf k}}}-
\frac{\lambda^{*}_{\rm so}({\bf k})T'_{a{\bf k}}}{\omega+E_{a{\bf k}}}\right],\nonumber\\
F_{2}^{\dagger}({\bf k},\omega)=\sum_{a=1,2}\left[\frac{\lambda_{\rm so}({\bf k})T_{a{\bf k}}}{\omega-E_{a{\bf k}}}-
\frac{\lambda_{\rm so}({\bf k})T_{a{\bf k}}}{\omega+E_{a{\bf k}}}\right],
\end{eqnarray}
where $a=1,2$ is used to denote four possible quasiparticle energy spectra, namely $\pm E_{1\bf{k}}$ and  $\pm E_{2\bf{k}}$ with
\begin{eqnarray}\label{Quasi-en-spec}
E_{1\bf{k}}&=&\sqrt{h^{2}+\xi^{2}_{\bf k}+\Delta_{so}^{2}({\bf k})+2\sqrt{h^{2}(\xi^{2}_{\bf k}+\Delta^{2})+\xi^{2}_{\bf k}\lambda^{2}_{\rm so}({\bf k})}},\nonumber\\
E_{2\bf{k}}&=&\sqrt{h^{2}+\xi^{2}_{\bf k}+\Delta_{so}^{2}({\bf k})-2\sqrt{h^{2}(\xi^{2}_{\bf k}+\Delta^{2})+\xi^{2}_{\bf k}\lambda^{2}_{\rm so}({\bf k})}}.\nonumber\\
\end{eqnarray}
Here $\Delta_{so}^{2}({\bf k})=\Delta^{2}+\lambda^{2}_{\rm so}({\bf k})$.  The expressions of all weight factors in above Green's functions $U'^2_{a{\bf k}}$, $V'^2_{a{\bf k}}$, $U^2_{a{\bf k}}$, $V^2_{a{\bf k}}$, $\alpha_{a{\bf k}}$,$\beta_{a{\bf k}}$, $P_{a{\bf k}}$, $Q_{a{\bf k}}$, $T'_{a{\bf k}}$ and $T_{a{\bf k}}$  are given in in Supplementary Material I.

Under the given Zeeman field strength and SOC strength, the chemical potential $\mu$ and pairing-gap order parameter $\Delta$ can be self-consistently determined by solving the density equation $n=n_{\uparrow}+n_{\downarrow}$ and order parameter equation $\Delta=U\langle \Psi_{\downarrow}\Psi_{\uparrow} \rangle$, namely
\begin{eqnarray}\label{twoequtions}
n&=&\sum_{{\bf k}a}\left[\left(U'^2_{a{\bf k}}+U^2_{a{\bf k}}-V'^2_{a{\bf k}}-V^2_{a{\bf k}}\right)f(E_{a\bf{k}})+V'^2_{a{\bf k}}+V^2_{a{\bf k}}\right],\nonumber\\
 1&=&\sum_{{\bf k}a}\left[\frac{\alpha_{a{\bf k}}-\beta_{a{\bf k}}}{2E_{{a\bf k}}}f(E_{a\bf{k}})+\frac{\beta_{a{\bf k}}}{2E_{{a\bf k}}}\right],
\end{eqnarray}
in which $n$ is the average occupancy number per lattice site and  $f(E_{a\bf{k}})$ is the Fermi distribution. We take $n=0.9$ in the whole paper. We have discussed the possible phase diagram of this system in our previous work \cite{Zhao2023-2}, the main results of which are that there are three distinct phase-separated regions in the ground phase diagram, including BCS superfluid (BCS-SF), topological superfluid (Topo-SF) and normal metal phases. These results can help to select a proper SOC strength and magnetic field to investigate the topological superfluid in this lattice system. For a given SOC strength $\lambda$,
the magnetic field $h$ is required to be bigger than a critical value to make system come into the topological superfluid, in which $\Delta$ first increases and then decreases as $\lambda$ increases. In particular, the $\lambda$ dependence of $\Delta$ in an optical lattice is different from the continuous Fermi gases where SOC can enhance the pairing gap strength monotonically \cite{Lee17,Hu11}.

All density related dynamical excitations can be investigated by the density dynamical structure factor, who is connected to the density response function $\chi_{D}=\chi_{11}+\chi_{22}+\chi_{12}+\chi_{21}$ by
\begin{equation} \label{DSF_RPA}
S({\bf q},{\omega})=-\frac{1}{\pi}\frac{1}{1-e^{-\omega/T}}{\rm Im}\chi_{D}({\bf q},i\omega_{n}\to \omega+i\delta),
\end{equation}
where $q$ and $\omega$ are the transferred momentum and energy, respectively. $\delta$ is a small positive number (usually we set $\delta=0.003$). The response function $\chi$ can be obtained by the random phase approximation (RPA). More details of $\chi$ under the RPA theory are presented in Supplementary Material II.

{\it Collective modes.}--Generally the full dynamical excitations consist of collective modes and single-particle excitations. By continuously increasing the Zeeman magnetic field over a critical value, a phase transition from a conventional BCS superfluid to a topological superfluid happens, which may bring some interesting characters to collective modes and single-particle excitations.

\begin{figure}
\includegraphics[scale=0.55]{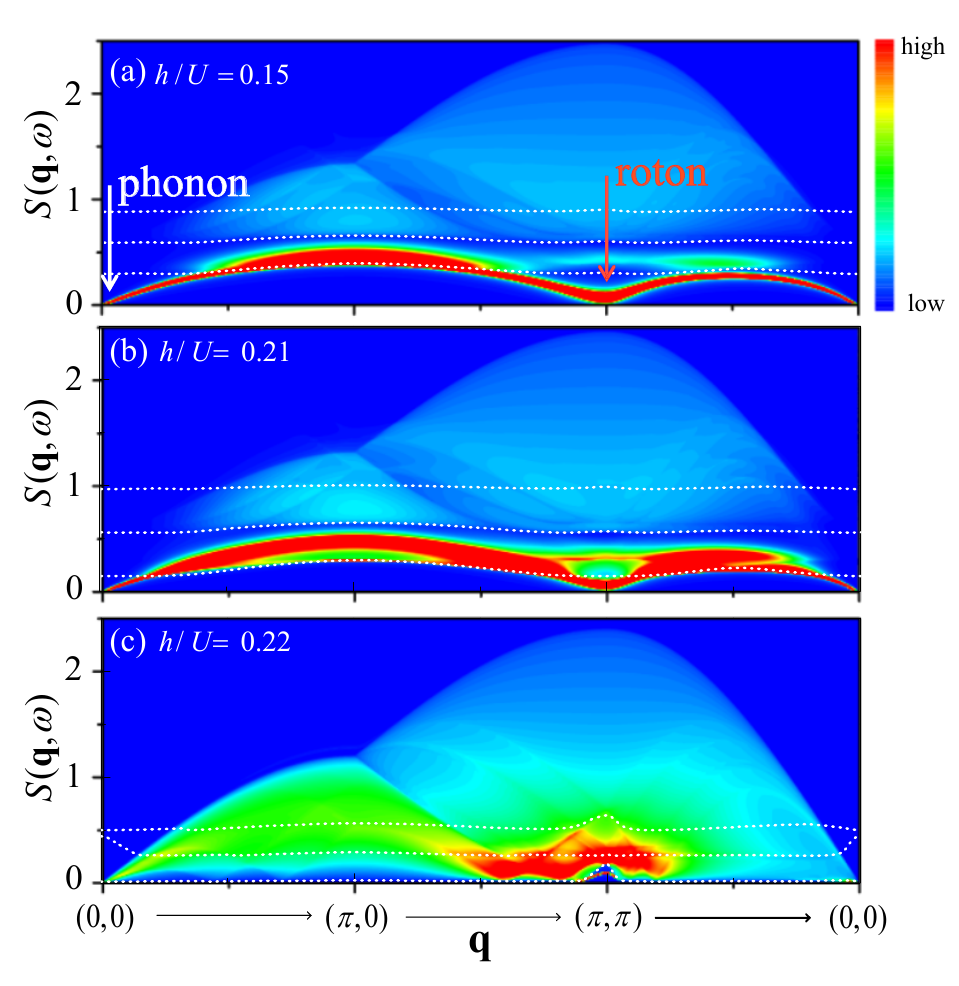}\caption{\label{fig1} Dynamical structure factor $S({\bf q},{\omega})$ along $[0,0]\rightarrow [\pi,0]\rightarrow [\pi,\pi]\rightarrow [0,0]$ in the BZ for (a) $h/U=0.15$ (BCS-SF), (b) $h/U=0.21$ (BCS-SF), and (c) $h/U=0.22$ (Topo-SF) with $\lambda/U=0.1$. The white dotted lines are the same lines as described
in Fig. \ref{lambh}, which show all kinds of threshold energy needed to break a Cooper pair.}
\end{figure}

 Here we discuss the external Zeeman field and SOC dependencies of all possible dynamical excitations, which are displayed by dynamical structure factor.  In Fig. \ref{fig1}, we numerically calculate the dynamical structure factor $S({\bf q},{\omega})$ along the high symmetry directions of the first Brillouin zone (BZ), namely $[0,0]\rightarrow [\pi,0]\rightarrow [\pi,\pi]\rightarrow [0,0]$,  at different external Zeeman fields. We take a Rashba SOC strength $\lambda/U=0.1$.
At the small transferred momentum region marked by the white arrow, $S({\bf q},{\omega})$ shows a sharp narrow peak ($\delta$-like) and a linear dispersion starting from zero energy appears. This linear dispersion is the collective phonon mode, and the slope of this dispersion at $\omega \rightarrow 0$ indicates the sound speed $c_{\rm s}=\omega/q$. The gapless collective phonon mode origins from the spontaneously U(1) phase symmetry breaking of order parameter in the superfluid state. The white dotted lines mark the location of three types of the minimum energy needed to break a Cooper pair, which will be introduced later. At the larger momentum, the collective phonon mode gradually merges into the single-particle excitations, and shows a finite expansion width because of the scattering with the single-particle excitations. In an optical lattice system, the continuous translational symmetry is also broken, which potentially induces another collective mode at $q=[\pi,\pi]$, which is called roton mode and is related to the global pseudospin SU(2) symmetry \cite{Zhang1990,Ganesh2009,Qin2022}. At half-filling ($\mu=0$), the roton mode is gapless, and there is a degeneracy between superfluid and charge density wave (CDW). Away from half-filling ($\mu \neq 0$), the roton mode is gapped which arises from the strong local density correlations \cite{Ganesh2009}. In an optical lattice system, this roton mode is always well separated from single-particle excitations. Obviously, the dynamical structure factor has a dramatic change from a BCS superfluid to a topological superfluid, which can be attributed to the discontinuous decrease of the pairing gap at the transition point.

\begin{figure*}[t]
\centering
\includegraphics[width=0.9\textwidth]{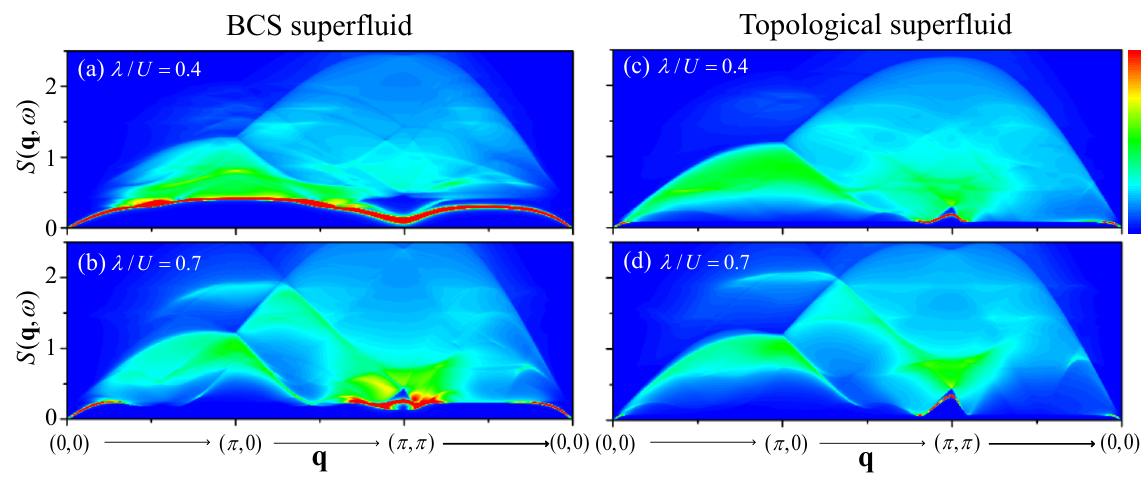}
\caption{\label{fig2} $S({\bf q},{\omega})$ along  $[0,0]\rightarrow [\pi,0]\rightarrow [\pi,\pi]\rightarrow [0,0]$ for SOC strength (a) (c) $\lambda/U=0.4$, (b) (d) $\lambda/U=0.7$ with the BCS superfluid $h/U=0.15$ (Left) and the topological superfluid $h/U=0.35$ (Right).
}
\end{figure*}

 In Fig. \ref{fig2}, we numerically calculate the SOC strength dependence of  $S({\bf q},{\omega})$ along the high symmetry directions under the BCS superfluid $h/U=0.15$ (Left) and the topological superfluid $h/U=0.35$ (Right), respectively.
At $\lambda/U=0.4$ in BCS superfluid, the collective mode is robust along the whole high symmetry directions. But when $\lambda/U=0.7$, the collective mode is suppressed owing to the scattering with the single-particle excitations and located around ${\bf q}=[0,0]$ and ${\bf q}=[\pi,\pi]$. When the system enters into the topological superfluid, the collective mode shrinks further.  Interestingly the collective roton mode at ${\bf q}=[\pi,\pi]$ is still robust, even though the collective phonon mode around ${\bf q}=[0,0]$ is weak. Also the excitation gap of roton mode increases with the Rashba SOC strength $\lambda$. We will find that the collective roton mode at ${\bf q}=[\pi,\pi]$ is the signal of molecular excitations, and it is closely related to the pairing gap.

\begin{figure}
\includegraphics[scale=0.48]{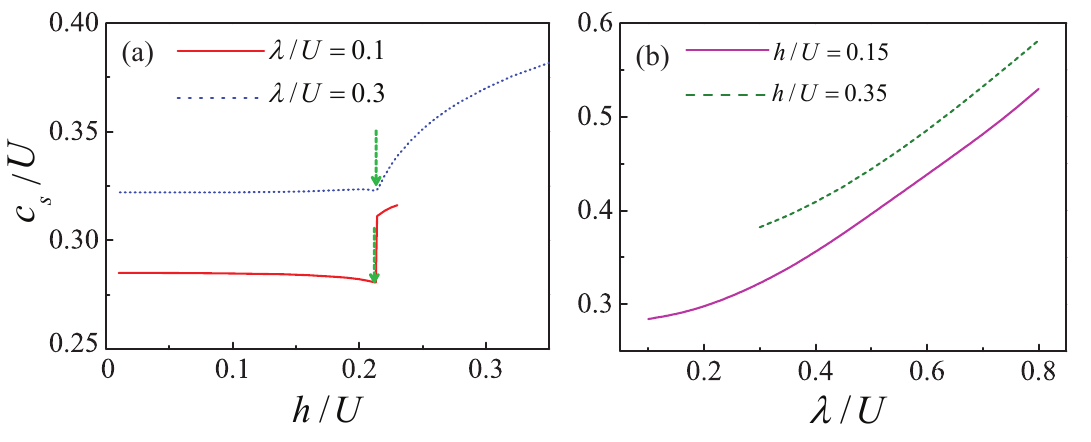}\caption{\label{fig3} (a) Sound speed $c_{s}$ as a function of the Zeeman field strength $h$ for the Rashba SOC strength $\lambda/U=0.1$ (red solid line), $\lambda/U=0.3$ (blue dotted line). (b) $c_{s}$ vs $\lambda$ for  $h/U=0.15$ (magenta solid line), $h/U=0.35$ (olive dashed line). The topological phase transition points are marked by the green arrows.}
\end{figure}
To show the main characters of collective phonon mode clearly, we calculate the $h$ and $\lambda$ dependencies of the sound speed, and the results are shown by two panels of Fig. \ref{fig3}.
From Fig. \ref{fig3}a, we find that the sound speed almost does not change with $h$ in the BCS superfluid, which can be explained by the Meissner effect. However, it suddenly rises when $h$ is over the critical Zeeman field $h_{c}$, over which the system comes into the topological superfluid state. Therefore, a minimum of the sound speed is located at the topological phase transition point, which is similar to the case in continuous Fermi gases \cite{He2013}. In particular, the sound speed has a discontinuous increase at a small SOC strength $\lambda/U=0.1$ where a first-order phase transition appears. However, when $\lambda/U=0.3$ (a second-order phase transition), the sound speed starts to increase steadily with $h$. Moreover, in Fig. \ref{fig3}b, the sound speed is proportional to $\lambda$ in both BCS superfluid and topological superfluid, which is contrast to the continuous Fermi gases \cite{He2013} but consistent with the numerical calculation results \cite{Koinov2017}.

{\it Single-particle excitations.}--In the high-energy region, dynamical excitations of the system are dominated by the single-particle excitations. The single-particle excitations show a very complex structure due to influence from the Zeeman field and SOC effect. Most of dynamical excitations are related to the pair-breaking effect. The two atoms in a Cooper pair can come from  quasiparticle spectra $E_{1\bf{k}}$ or $E_{2\bf{k}}$, which leads to the appearance of four kinds of mechanism to break Cooper pairs, which are respectively $\{11\}\rightarrow E_{1\bf{k+q}}+E_{1\bf{k}}$, $\{12\}\rightarrow E_{1\bf{k+q}}+E_{2\bf{k}}$, $\{21\}\rightarrow E_{2\bf{k+q}}+E_{1\bf{k}}$, and $\{22\}\rightarrow E_{2\bf{k+q}}+E_{2\bf{k}}$. Usually the $\{12\}$- and $\{21\}$-type excitations are overlapped with each other, so only three kinds of pair-breaking excitations display. More details had been shown in Supplementary Material III.

\begin{figure}
\includegraphics[scale=0.45]{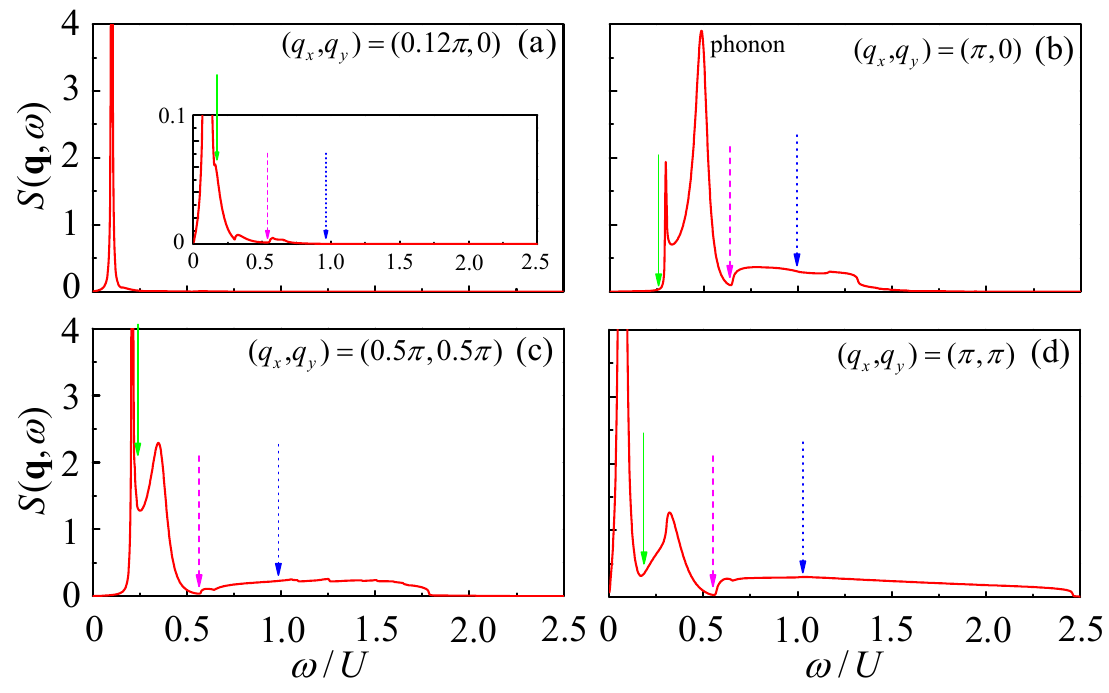}\caption{\label{fig4} Dynamical structure factor $S({\bf q},{\omega})$ of the BCS superfluid with Zeeman field $h/U=0.21$ at different transferred momenta: $[0.12\pi,0]$,  (b)$[\pi,0]$, (c) $[0.5\pi,0.5\pi]$, and (d) $[\pi,\pi]$ with $\lambda/U=0.1$. The green solid, magenta dashed and blue dotted arrows respectively mark the threshold energy in the $\{11\}-$, $\{12\}-$, and $\{22\}-$type excitations.}
\end{figure}

To clearly show the energy and momentum dependencies of dynamical excitations, we calculate the energy dependence of $S({\bf q},{\omega})$ at a few typical transferred momenta in the BZ. The related results obtained in BCS superfluid $h/U=0.21$ as a function of ${\omega}$ with $\lambda/U=0.1$ for different transferred momenta are plotted in Fig. \ref{fig4}. The green solid, magenta dashed and blue dotted arrows respectively mark the minimum energy to break a Cooper pair in the $\{11\}$-, $\{12\}$ (or $\{21\}$)-, and $\{22\}$-type excitations.
In panel (a), a strong sharp phonon peak appears at the low-energy region, and three kinds of the single-particle excitations appearing at the larger energy are weak. When ${\bf q}=[\pi,0]$ in panel (b), owing to the SOC effect, the $\{22\}-$type pair-break excitation happens at the smaller energy than the phonon peak.  So the phonon peak is suppressed and has a finite width because the phonon mode is pushed up to the single-particle excitation region and competes with them. In panel (c), the phonon peak appears earlier than the $\{22\}-$type pair-break excitation, which gives a strong sharp phonon peak. In panel (d), the sharp collective mode corresponds to the molecular excitations, which is found to be sensitive to variations in the pairing gap. Recent Bragg experiments have also found this feature in Fermi gases (arXiv:2310.03452).

{\it Measuring pairing gap at ${\bf q}=[\pi,\pi]$.}--Due to continuous translational symmetry breaking of the lattice system, an obvious roton collective mode tends to display at the specific transferred momentum ${\bf q}=[\pi,\pi]$, which is just the boundary of BZ. The roton mode is called the molecular Cooper-pair excitations here \cite{Zhang1990}, and it is interesting to note that this mode is well separated from the continuous pair-breaking excitations. This provides a potential way to experimentally measure the pairing gap $\Delta$, which has been introduced in our previous work k (arXiv:2305.09685). It is interesting to check that whether the same strategy can also work in the SOC lattice system, or not.

\begin{figure*}[t]
\centering
\includegraphics[width=0.9\textwidth]{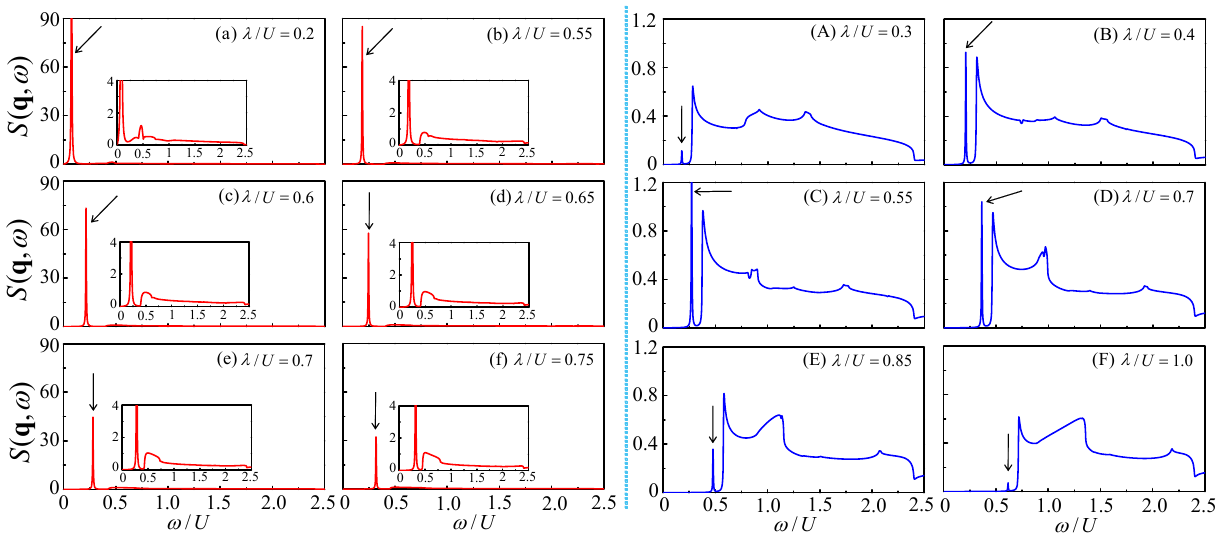}
\caption{\label{BCStopo_lambda}  Left two columns: $S({\bf q}=[\pi,\pi],{\omega})$ of BCS superfluid with $h/U=0.2$ for different SOC strength (a) $\lambda/U=0.2$, (b) $\lambda/U=0.55$, (c) $\lambda/U=0.6$, (d) $\lambda/U=0.65$, (e) $\lambda/U=0.7$, (f) $\lambda/U=0.75$. Right two columns: $S({\bf q}=[\pi,\pi],{\omega})$ of the topological superfluid with $h/U=0.45$ for (A) $\lambda/U=0.3$, (B) $\lambda/U=0.4$, (C) $\lambda/U=0.55$, (D) $\lambda/U=0.7$, (E) $\lambda/U=0.85$, (F) $\lambda/U=1.0$. The arrow marks the location of the molecule excitation peak..
}
\end{figure*}
To study the relationship between the molecular excitation (roton mode marked by the red arrow in Fig. {\ref{fig1}}a) and pairing gap $\Delta$, we study the Zeeman field and SOC strength dependencies of dynamical structure factor. As shown in Fig. \ref{BCStopo_lambda}, we investigate the dynamical structure factor $S({\bf q}=[\pi,\pi],{\omega})$ in both BCS superfluid and topological superfluid. Here the dynamical structure factor of two figures consists of a sharp Cooper-pair molecular excitation peak (marked by the arrow) and a broad continuous single-particle excitation regime. In left panel of Fig. \ref{BCStopo_lambda} (BCS superfluid), it is obvious to see that the height of the molecules peak gradually decreases for a more and more large SOC strength $\lambda$. However, a non-monotonous behavior of the height of the molecular peak in the topological superfluid (right panel of Fig. \ref{BCStopo_lambda}) is displayed. The height is initially proportional to $\lambda$ in the low $\lambda$ region, and then reaches a maximum height around $\lambda/U\approx 0.55$, and finally decreases with $\lambda$. It is interesting to note that the $\lambda$-dependence behaviour of molecular peak is consistent with $\lambda$ dependence of the pairing gap, even with the influence from the Zeeman field and SOC effect.

\begin{figure}
\includegraphics[scale=0.45]{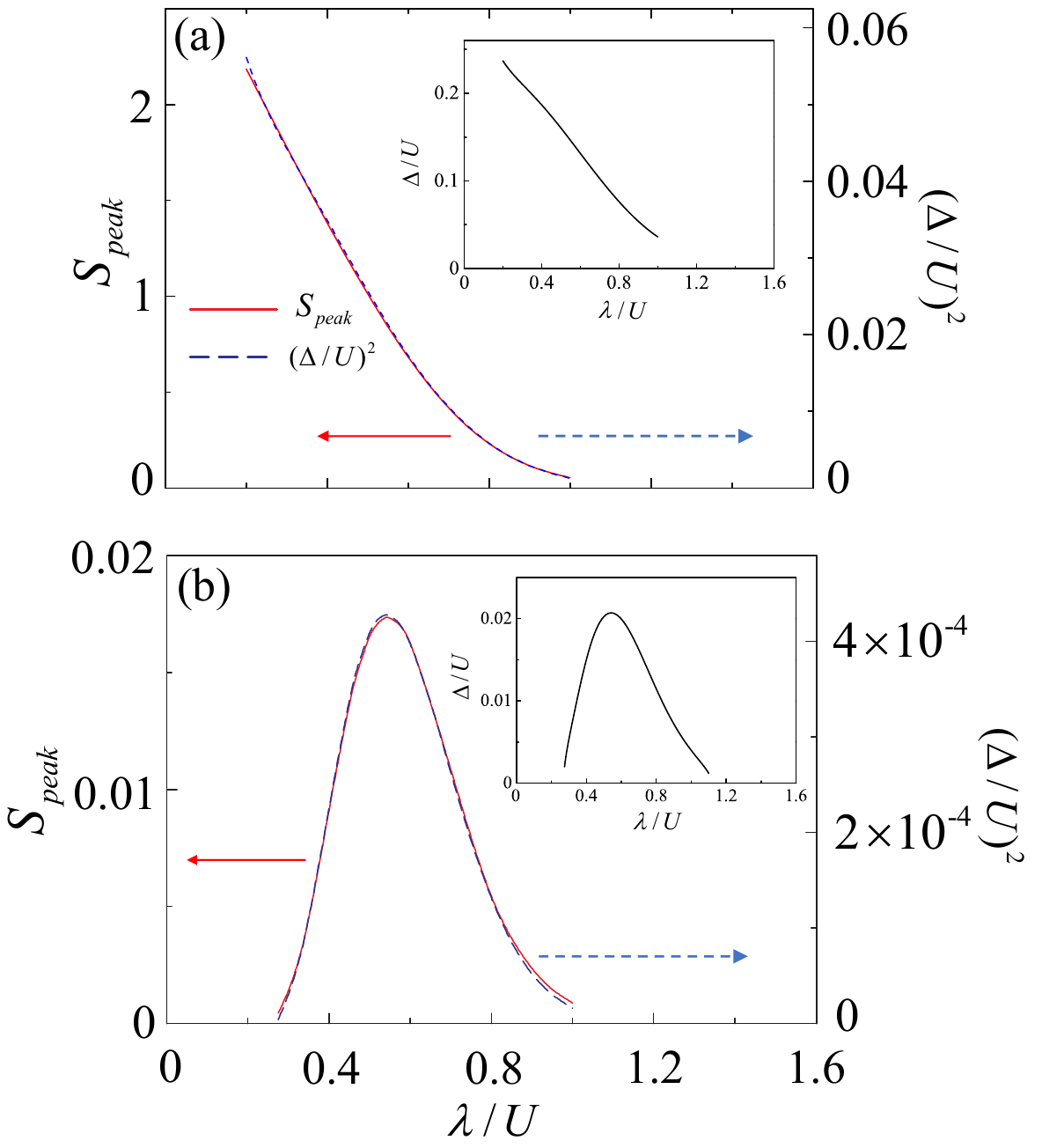}\caption{\label{weight} The relation between covering area of the molecular peak (red solid line) and SOC strength $\lambda$, compared with the square of pairing gap $\Delta^{2}$ (blue dashed line), in both (a) BCS superfluid ($h/U=0.2$) and (b) topological superfluid ($h/U=0.45$). Inset: the corresponding pairing gap $\Delta$ as a function of $\lambda$.}
\end{figure}

To demonstrate the connection between the molecular excitations and pairing gap clearly, we extract the covered area $S_{peak}$ of the molecular peak from Fig. \ref{BCStopo_lambda}, and plot $S_{peak}$ (red solid line) as a function of SOC strength $\lambda$ in Fig. \ref{weight}, in comparison with the pairing density of Cooper pairs (namely the square of pairing gap $\Delta^{2}$ (blue dashed line)).
Our results show that both $S_{peak}$ and $\Delta^{2}$ display an almost the same tendency for different $\lambda$. Moreover, we also discuss the Zeeman field dependence of both $S_{peak}$ and $\Delta^2$ in Fig. \ref{hdepen}. These results verify the conclusions that the weight of molecular excitation peak (namely $S_{peak}$) is linearly determined by the square of pairing gap $\Delta^2$. The similar physics had been introduced in superconductors by S.-C. Zhang \cite{Zhang1990}. Also the molecular excitations is well separated from the continuous pair-breaking excitation. These results provide an important inspire to us that experimentally one can directly measure the magnitude of the pairing gap by detecting the area of the molecular excitation peak at ${\bf q}=[\pi,\pi]$ in the lattice system. Much more significantly, this strategy is universal in the lattice system, no matter the existence of the SOC effect and Zeeman field.

{\it Conclusion.}---Dynamical excitations of the Fermi gases in a 2D optical lattice with Rashba spin-orbit coupling are discussed within RPA theory. By numerically calculating dynamical structure factor, the collective phonon mode can be obtained in a small transferred momentum region and the collective roton mode appears at ${\bf q}=[\pi,\pi]$. Our results show that the sound speed has different Zeeman field $h$-dependence behaviour between the BCS superfluid and topological superfluid. As $h$ increases, the sound speed is almost constant in the BCS superfluid, but continuously increases when the system enters into the topological state. Moreover, the value of sound speed is determined to $\lambda$ in both the BCS superfluid and the topological superfluid. By investing the dynamical excitation at ${\bf q}=[\pi,\pi]$, we find that the area of the molecular excitation peak scales with the square of pairing gap, which indicates that experimentally the pairing gap can be measured by detecting the molecular excitation peak at ${\bf q}=[\pi, \pi]$ in the BZ.

 Recently an experimental scheme to realize the 2D attractive Fermi-Hubbard model with SOC was proposed in ultracold fermions by Peking University (arXiv:2309.12923). This scheme is essential in realizing the topological superfluidity for the first time. Therefore, it is also important for us to study the physical properties of the attractive Fermi-Hubbard model with SOC.

{\it Author contributions.}---
HZ conceived the project. The model, RPA theory and physical explanation were finished by HZ, FY, and PZ. The fortran codes and most figures were finished by HZ, and RH. The sound speed was calculated by LQ. HZ and PZ wrote the manuscript with inputs from all coauthors.

{\it Conflict of interest.}---
The authors declare that they have no conflict of interest.

{\it Acknowledgments.}---
This research was supported by the National Natural Science Foundation of China,
Grants No. 11547034 (H.Z.), No. 11804177 (P.Z.).

\newpage
\twocolumngrid

%%%%%%%%%%%%%%%%%%%%%%%%%%%%%%%%%%%%%%
%%   Supplementary Information
%%%%%%%%%%%%%%%%%%%%%%%%%%%%%%%%%%%%%%
%\appendix
\renewcommand{\thesection}{S-\arabic{section}}
\setcounter{section}{0}  %  this will re-count section from 1
\renewcommand{\theequation}{S\arabic{equation}}
\setcounter{equation}{0}  %  this will re-count eq from 1
\renewcommand{\thefigure}{S\arabic{figure}}
\setcounter{figure}{0}  %  this will re-count eq from 1

\indent

In this supplementary material we provide the details of the weight factors of the Green's functions, the response functions, three pair-breaking excitations, and the Zeeman field dependence of both $S_{peak}$ and $\Delta^2$, which supports our main idea to measure the pairing gap through the dynamical structure factor.

%\settocdepth{section}
\addtocontents{toc}{\protect\setcounter{tocdepth}{2}}
\section{I. Weight factors}
\label{SI:weight}
 The weight factors are as follows:
\begin{eqnarray}\label{weightfactor}
 U'^2_{a{\bf k}}&=&(\Omega'_{a{\bf k}}+\Xi'_{a{\bf k}})/\Sigma_{a{\bf k}}, V'^2_{a{\bf k}}=(\Omega'_{a{\bf k}}-\Xi'_{a{\bf k}})/\Sigma_{a{\bf k}}\nonumber\\
 U^2_{a{\bf k}}&=&(\Omega_{a{\bf k}}+\Xi_{a{\bf k}})/\Sigma_{a{\bf k}}, V^2_{a{\bf k}}=(\Omega_{a{\bf k}}-\Xi_{a{\bf k}})/\Sigma_{a{\bf k}}\nonumber\\
 \alpha_{a{\bf k}}&=&[E^{2}_{a{\bf k}}+2hE_{a{\bf k}}+h^{2}-\xi^{2}_{\bf k}-\Delta_{so}^{2}({\bf k})]\Delta/\Sigma_{a{\bf k}},\nonumber\\
 \beta_{a{\bf k}}&=&-[E^{2}_{a{\bf k}}-2hE_{a{\bf k}}+h^{2}-\xi^{2}_{\bf k}-\Delta_{so}^{2}({\bf k})]\Delta/\Sigma_{a{\bf k}},\nonumber\\
 P_{a{\bf k}}&=&[E^{2}_{a{\bf k}}+2\xi_{\bf k}E_{a{\bf k}}+\xi^{2}_{\bf k}-h^{2}-\Delta_{so}^{2}({\bf k})]/\Sigma_{a{\bf k}},\nonumber\\
 Q_{a{\bf k}}&=&-[E^{2}_{a{\bf k}}-2\xi_{\bf k}E_{a{\bf k}}+\xi^{2}_{\bf k}-h^{2}-\Delta_{so}^{2}({\bf k})]/\Sigma_{a{\bf k}}, \nonumber
\end{eqnarray}
where the functions are given as:
\begin{eqnarray}\label{kernalfunction}
 \Omega'_{a{\bf k}}&=&E^{3}_{a{\bf k}}-[(\xi_{\bf k}+h)^{2}+\Delta_{so}^{2}({\bf k})]E_{a{\bf k}},\nonumber\\
 \Xi'_{a{\bf k}}&=&(\xi_{\bf k}-h)E^{2}_{a{\bf k}}-(\xi_{\bf k}+h)[\xi^{2}_{\bf k}+\Delta^{2}-h^{2}-\lambda^{2}_{\rm so}({\bf k})],\nonumber\\
 \Omega_{a{\bf k}}&=&E^{3}_{a{\bf k}}-[(\xi_{\bf k}-h)^{2}+\Delta_{so}^{2}({\bf k})]E_{a{\bf k}},\nonumber\\
 \Xi_{a{\bf k}}&=&(\xi_{\bf k}+h)E^{2}_{a{\bf k}}-(\xi_{\bf k}-h)[\xi^{2}_{\bf k}+\Delta^{2}-h^{2}-\lambda^{2}_{\rm so}({\bf k})],\nonumber\\
 \Sigma_{1{\bf k}}&=&2(E^{2}_{1{\bf k}}-E^{2}_{2{\bf k}})E_{1{\bf k}}, \Sigma_{2{\bf k}}=2(E^{2}_{2{\bf k}}-E^{2}_{1{\bf k}})E_{2{\bf k}}\nonumber\\
 T'_{a{\bf k}}&=&2(h+\xi_{\bf k})\Delta/\Sigma_{a{\bf k}}, T_{a{\bf k}}=2(h-\xi_{\bf k})\Delta/\Sigma_{a{\bf k}}. \nonumber
\end{eqnarray}
The weight factors satisfy the sum rule: $\sum_{a}(U^2_{a{\bf k}}+V^2_{a{\bf k}})=1$,  $[\sum_{a}(U'^2_{a{\bf k}}+V'^2_{a{\bf k}})=1]$.

\section{II. Random phase approximation and Response functions}
\label{SI:response}
In a Fermi superfluid, there are four different densities, including the spin-up density
$n_{1}=\left\langle \psi_{\uparrow}^{\dagger}\psi_{\uparrow}\right\rangle $, the spin-down density
$n_{2}=\left\langle \psi_{\downarrow}^{\dagger}\psi_{\downarrow}\right\rangle $, the anomalous pairing density
$n_{3}=\left\langle \psi_{\downarrow}\psi_{\uparrow}\right\rangle $ and
its conjugate part $n_{4}=\left\langle \psi_{\uparrow}^{\dagger}\psi_{\downarrow}^{\dagger}\right\rangle $.
The interaction makes these densities coupled with each other. Any fluctuation in each kind of density will influence other densities and generate an obvious density fluctuation of them. Also any weak perturbation potential $V_{\textrm{pert}}$ will generate density fluctuations $\delta n$, and they are connected with each other by response function $\chi$, namely $\delta n=\chi V_{\textrm{pert}}$, in the frame of linear response theory. In the mean-field level, the mean-field response
function is a $4\times4$ matrix
\begin{eqnarray}
\chi^{0}=\left[\begin{array}{cccc}
\chi_{11}^{0} & \chi_{12}^{0} & \chi_{13}^{0} & \chi_{14}^{0}\\
\chi_{21}^{0} & \chi_{22}^{0} & \chi_{23}^{0} & \chi_{24}^{0}\\
\chi_{31}^{0} & \chi_{32}^{0} & \chi_{33}^{0} & \chi_{34}^{0}\\
\chi_{41}^{0} & \chi_{42}^{0} & \chi_{43}^{0} & \chi_{44}^{0}
\end{array}\right]
\end{eqnarray}
with $\chi_{ij}^{0}\left(r_{1},r_{2},\tau,0\right)=-\left\langle \hat{n}_{i}\left(r_{1},\tau\right)\hat{n}_{j}\left(r_{2},0\right)\right\rangle $ in coordinate $r$ and imaginary time $\tau$ space. These 16 matrix elements are determined by the corresponding density-density correlation functions which can be obtained by defining corresponding Green's functions. In fact, as a result of the symmetry of system, only 10 matrix elements are independent, i.e., $\chi^{0}_{12}=\chi^{0}_{21}$, $\chi^{0}_{33}=\chi^{0}_{44}$,
 $\chi^{0}_{31}=\chi^{0}_{14}$, $\chi^{0}_{32}=\chi^{0}_{24}$,
 $\chi^{0}_{41}=\chi^{0}_{13}$, $\chi^{0}_{42}=\chi^{0}_{23}$.
 Based on the mean-field theory by calculating the corresponding Green's functions in 2D Fermi gas with Rashba-type SOC, we can obtain the correlation functions as:
\begin{eqnarray}\label{a11}
 \chi^{0}_{11}&=&\sum_{kaa'}\left(U'^2_{a{k}}U'^2_{a'{k+q}}-\lambda^{2}_{R}T'_{a{k}}T'_{a'{k+q}}\right)I_{1}\left(k,q,i\omega_{n}\right)\nonumber\\
&+&\sum_{kaa'}\left(U'^2_{a{k}}V'^2_{a'{k+q}}+\lambda^{2}_{R}T'_{a{k}}T'_{a'{k+q}}\right)I_{2}\left(k,q,i\omega_{n}\right)\nonumber\\
&+&\sum_{kaa'}\left(V'^2_{a{k}}U'^2_{a'{k+q}}+\lambda^{2}_{R}T'_{a{k}}T'_{a'{k+q}}\right)I_{3}\left(k,q,i\omega_{n}\right)\nonumber\\
 &+&\sum_{kaa'}\left(V'^2_{a{k}}V'^2_{a'{k+q}}-\lambda^{2}_{R}T'_{a{k}}T'_{a'{k+q}}\right)I_{4}\left(k,q,i\omega_{n}\right), \nonumber
\end{eqnarray}
\begin{eqnarray}\label{b12}
\chi^{0}_{12}=&-&\sum_{kaa'}\left(\alpha_{a{k}}\alpha_{a'{k+q}}-\lambda^{2}_{R}P_{a{k}}P_{a'{k+q}}\right)I_{1}\left(k,q,i\omega_{n}\right)\nonumber\\
 &+&\sum_{kaa'}\left(\alpha_{a{k}}\beta_{a'{k+q}}-\lambda^{2}_{R}P_{a{k}}Q_{a'{k+q}}\right)I_{2}\left(k,q,i\omega_{n}\right)\nonumber\\
 &+&\sum_{kaa'}\left(\beta_{a{k}}\alpha_{a'{k+q}}-\lambda^{2}_{R}Q_{a{k}}P_{a'{k+q}}\right)I_{3}\left(k,q,i\omega_{n}\right)\nonumber\\
 &+&\sum_{kaa'}\left(\beta_{a{k}}\beta_{a'{k+q}}-\lambda^{2}_{R}Q_{a{k}}Q_{a'{k+q}}\right)I_{4}\left(k,q,i\omega_{n}\right), \nonumber
\end{eqnarray}
\begin{eqnarray}\label{a22}
\chi^{0}_{22}&=&\sum_{kaa'}\left(U^2_{a{k}}U^2_{a'{k+q}}-\lambda^{2}_{R}T_{a{k}}T_{a'{k+q}}\right)I_{1}\left(k,q,i\omega_{n}\right)\nonumber\\
 &+&\sum_{kaa'}\left(U^2_{a{k}}V^2_{a'{k+q}}+\lambda^{2}_{R}T_{a{k}}T_{a'{k+q}}\right)I_{2}\left(k,q,i\omega_{n}\right)\nonumber\\
 &+&\sum_{kaa'}\left(V^2_{a{k}}U^2_{a'{k+q}}+\lambda^{2}_{R}T_{a{k}}T_{a'{k+q}}\right)I_{3}\left(k,q,i\omega_{n}\right)\nonumber\\
 &+&\sum_{kaa'}\left(V^2_{a{k}}V^2_{a'{k+q}}-\lambda^{2}_{R}T_{a{k}}T_{a'{k+q}}\right)I_{4}\left(k,q,i\omega_{n}\right), \nonumber
\end{eqnarray}
\begin{eqnarray}\label{c13}
\chi^{0}_{13}&=&\sum_{kaa'}(U'^2_{a'{k}}\beta_{a{k+q}}-\lambda^{2}_{R}T'_{a{k+q}}P_{a'{k}})I_{1}(k,q,i\omega_{n})\nonumber\\
 &+&\sum_{kaa'}(U'^2_{a'{k}}\alpha_{a{k+q}}+\lambda^{2}_{R}T'_{a{k+q}}P_{a'{k}})I_{2}(k,q,i\omega_{n})\nonumber\\
 &+&\sum_{kaa'}(V'^2_{a'{k}}\beta_{a{k+q}}-\lambda^{2}_{R}T'_{a{k+q}}Q_{a'{k}})I_{3}(k,q,i\omega_{n})\nonumber\\
 &+&\sum_{kaa'}(V'^2_{a'{k}}\alpha_{a{k+q}}+\lambda^{2}_{R}T'_{a{k+q}}Q_{a'{k}})I_{4}(k,q,i\omega_{n}), \nonumber
\end{eqnarray}
\begin{eqnarray}\label{c14}
\chi^{0}_{14}&=&\sum_{kaa'}(U'^2_{a{k+q}}\beta_{a'{k}}-\lambda^{2}_{R}T'_{a'{k}}P_{a{k+q}})I_{1}(k,q,i\omega_{n})\nonumber\\
 &+&\sum_{kaa'}(V'^2_{a{k+q}}\beta_{a'{k}}-\lambda^{2}_{R}T'_{a'{k}}Q_{a{k+q}})I_{2}(k,q,i\omega_{n})\nonumber\\
 &+&\sum_{kaa'}(U'^2_{a{k+q}}\alpha_{a'{k}}+\lambda^{2}_{R}T'_{a'{k}}P_{a{k+q}})I_{3}(k,q,i\omega_{n})\nonumber\\
 &+&\sum_{kaa'}(V'^2_{a{k+q}}\alpha_{a'{k}}+\lambda^{2}_{R}T'_{a'{k}}Q_{a{k+q}})I_{4}(k,q,i\omega_{n}), \nonumber
\end{eqnarray}
\begin{eqnarray}\label{c23}
\chi^{0}_{23}=&-&\sum_{kaa'}(U^2_{a'{k}}\alpha_{a{k+q}}-\lambda^{2}_{R}P_{a'{k}}T_{a{k+q}})I_{1}(k,q,i\omega_{n})\nonumber\\
 &+&\sum_{kaa'}(U^2_{a'{k}}\beta_{a{k+q}}+\lambda^{2}_{R}P_{a'{k}}T_{a{k+q}})I_{2}(k,q,i\omega_{n})\nonumber\\
 &+&\sum_{kaa'}(V^2_{a'{k}}\alpha_{a{k+q}}-\lambda^{2}_{R}Q_{a'{k}}T_{a{k+q}})I_{3}(k,q,i\omega_{n})\nonumber\\
 &+&\sum_{kaa'}(V^2_{a'{k}}\beta_{a{k+q}}+\lambda^{2}_{R}Q_{a'{k}}T_{a{k+q}})I_{4}(k,q,i\omega_{n}), \nonumber
\end{eqnarray}
\begin{eqnarray}\label{c24}
\chi^{0}_{24}=&-&\sum_{kaa'}(U^2_{a{k+q}}\alpha_{a'{k}}-\lambda^{2}_{R}T_{a'{k}}P_{a{k+q}})I_{1}(k,q,i\omega_{n})\nonumber\\
 &+&\sum_{kaa'}(V^2_{a{k+q}}\alpha_{a'{k}}-\lambda^{2}_{R}T_{a'{k}}Q_{a{k+q}})I_{2}(k,q,i\omega_{n})\nonumber\\
 &+&\sum_{kaa'}(U^2_{a{k+q}}\beta_{a'{k}}+\lambda^{2}_{R}T_{a'{k}}P_{a{k+q}})I_{3}(k,q,i\omega_{n})\nonumber\\
 &+&\sum_{kaa'}(V^2_{a{k+q}}\beta_{a'{k}}+\lambda^{2}_{R}T_{a'{k}}Q_{a{k+q}})I_{4}(k,q,i\omega_{n}), \nonumber
\end{eqnarray}
\begin{eqnarray}\label{m33}
\chi^{0}_{33}&=&\sum_{kaa'}(\beta_{a'{k}}\beta_{a{k+q}}-\lambda^{2}_{R}T_{a'{k}}T'_{a{k+q}})I_{1}(k,q,i\omega_{n})\nonumber\\
 &+&\sum_{kaa'}(\beta_{a'{k}}\alpha_{a{k+q}}+\lambda^{2}_{R}T_{a'{k}}T'_{a{k+q}})I_{2}(k,q,i\omega_{n})\nonumber\\
 &+&\sum_{kaa'}(\alpha_{a'{k}}\beta_{a{k+q}}+\lambda^{2}_{R}T_{a'{k}}T'_{a{k+q}})I_{3}(k,q,i\omega_{n})\nonumber\\
 &+&\sum_{kaa'}(\alpha_{a'{k}}\alpha_{a{k+q}}-\lambda^{2}_{R}T_{a'{k}}T'_{a{k+q}})I_{4}(k,q,i\omega_{n}), \nonumber
\end{eqnarray}
\begin{eqnarray}\label{ht34}
\chi^{0}_{34}&=&\sum_{kaa'}(V^2_{a'{k}}U'^2_{a{k+q}}+\lambda^{2}_{R}Q_{a'{k}}P_{a{k+q}})I_{1}(k,q,i\omega_{n})\nonumber\\
 &+&\sum_{kaa'}(V^2_{a'{k}}V'^2_{a{k+q}}+\lambda^{2}_{R}Q_{a'{k}}Q_{a{k+q}})I_{2}(k,q,i\omega_{n})\nonumber\\
 &+&\sum_{kaa'}(U^2_{a'{k}}U'^2_{a{k+q}}+\lambda^{2}_{R}P_{a'{k}}P_{a{k+q}})I_{3}(k,q,i\omega_{n})\nonumber\\
 &+&\sum_{kaa'}(U^2_{a'{k}}V'^2_{a{k+q}}+\lambda^{2}_{R}P_{a'{k}}Q_{a{k+q}})I_{4}(k,q,i\omega_{n}), \nonumber
\end{eqnarray}
\begin{eqnarray}\label{ht43}
\chi^{0}_{43}&=&\sum_{kaa'}(U^2_{a'{k}}V'^2_{a{k+q}}+\lambda^{2}_{R}P_{a'{k}}Q_{a{k+q}})I_{1}(k,q,i\omega_{n})\nonumber\\
 &+&\sum_{kaa'}(U^2_{a'{k}}U'^2_{a{k+q}}+\lambda^{2}_{R}P_{a'{k}}P_{a{k+q}})I_{2}(k,q,i\omega_{n})\nonumber\\
 &+&\sum_{kaa'}(V^2_{a'{k}}V'^2_{a{k+q}}+\lambda^{2}_{R}Q_{a'{k}}Q_{a{k+q}})I_{3}(k,q,i\omega_{n})\nonumber\\
 &+&\sum_{kaa'}(V^2_{a'{k}}U'^2_{a{k+q}}+\lambda^{2}_{R}Q_{a'{k}}P_{a{k+q}})I_{4}(k,q,i\omega_{n}), \nonumber
\end{eqnarray}
where $\lambda^{2}_{R}=\lambda^{2}(\sin^{2}{k_{x}}\cos{q_{x}}+\sin^{2}{k_{y}}\cos{q_{y}})$, the corresponding functions  $I_{1}(k,q,i\omega_{n})$,  $I_{2}(k,q,i\omega_{n})$, $I_{3}(k,q,i\omega_{n})$ and $I_{4}(k,q,i\omega_{n})$ are shown as
\begin{eqnarray}\label{Ikq}
 I_{1}(k,q,i\omega_{n})&=&\frac{f(E_{ak})-f(E_{a'k+q})}{i\omega_{n}+E_{ak}-E_{a'k+q}}\nonumber\\
 I_{2}(k,q,i\omega_{n})&=&\frac{f(E_{ak})+f(E_{a'k+q})-1}{i\omega_{n}+E_{ak}+E_{a'k+q}}\nonumber\\
 I_{3}(k,q,i\omega_{n})&=&\frac{1-f(E_{ak})-f(E_{a'k+q})}{i\omega_{n}-E_{ak}-E_{a'k+q}}\nonumber\\
 I_{4}(k,q,i\omega_{n})&=&\frac{f(E_{a'k+q})-f(E_{ak})}{i\omega_{n}-E_{ak}+E_{a'k+q}}.\nonumber
\end{eqnarray}
with $a(a')=1,2$.

 However, this mean-field response function fails to provide a qualitatively correct prediction of many dynamical excitations since it neglects the contribution from the fluctuation of interaction Hamiltonian. The random phase approximation picks back this fluctuation and treats it as part of an effective perturbation potential. Then one can find connection between the response function $\chi$ and the mean-field response function $\chi^0$, and this relation is given below
\begin{eqnarray}
\chi({\bf q},i\omega_{n})=\frac{\chi^{0}({\bf q},i\omega_{n})}{\hat{1}-\chi^{0}({\bf q},i\omega_{n})M_{I}U},
\end{eqnarray}
where $M_{I}=\sigma_{0}\otimes\sigma_{x}$ is a direct product of two Pauli matrices $\sigma_{0}$ and $\sigma_{x}$, and the unit matrix $\hat{1}=\sigma_{0}\otimes\sigma_{0}$.

\section{III. Minimum energy of three pair-breaking excitations}
\label{SI:pair-break}
\begin{figure}
\includegraphics[scale=0.4]{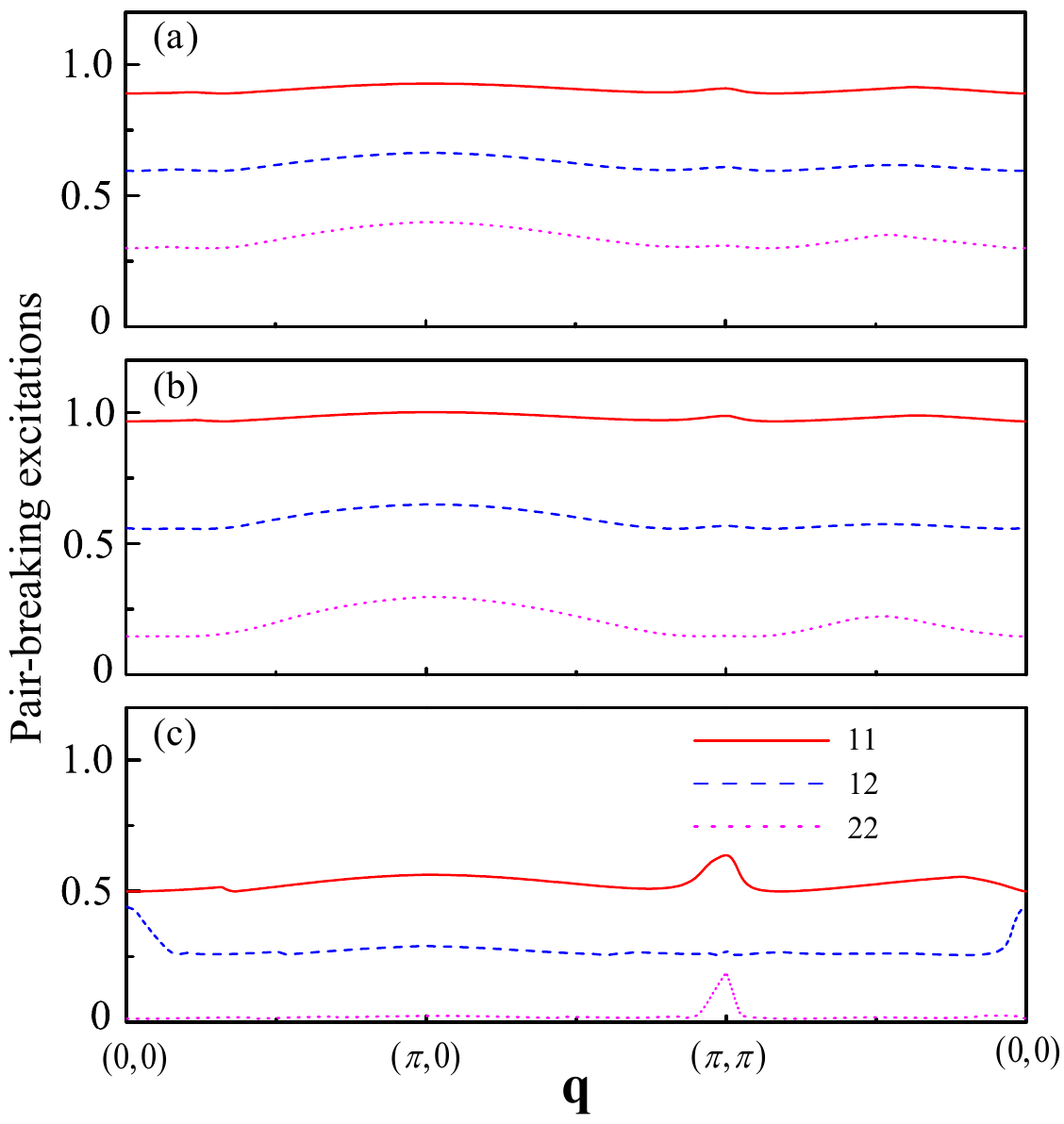}\caption{\label{lambh} All kinds of threshold energy of pair-breaking excitations
in different momenta for (a) $h/U=0.15$ (BCS-SF), (b) $h/U=0.21$ (BCS-SF), and (c) $h/U=0.22$ (Topo-SF) with $\lambda=0.1$.}
\end{figure}

\begin{figure}
\includegraphics[scale=0.45]{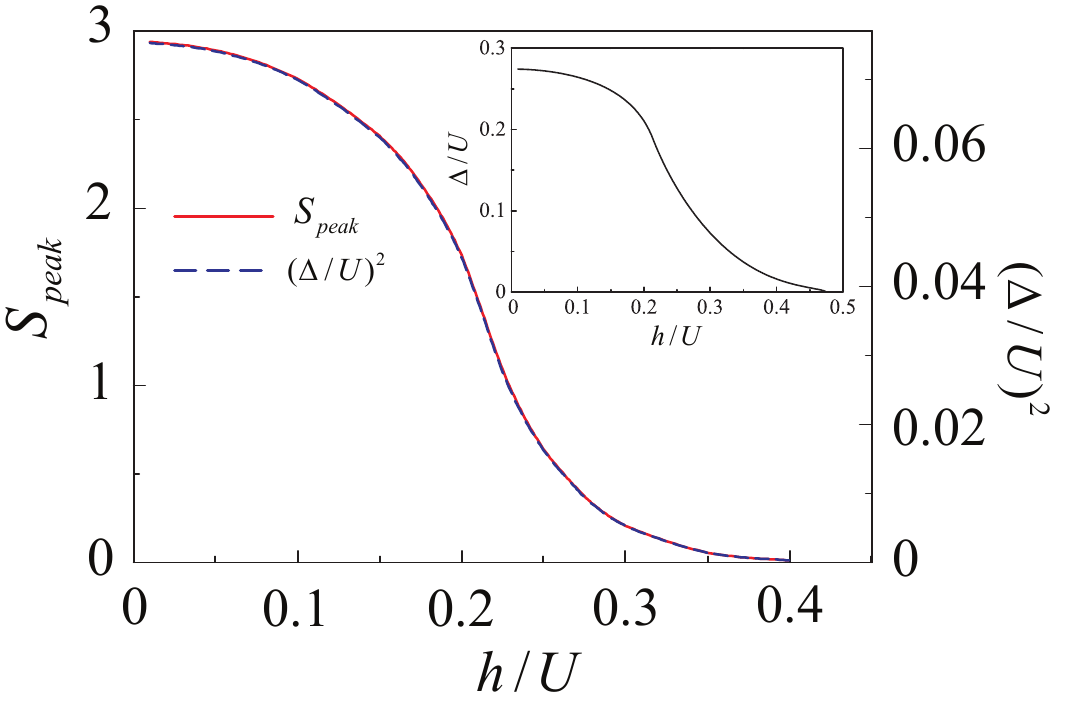}\caption{\label{hdepen} The relations between covering area of the molecular peak (red solid line) and Zeeman field $h$ for $\lambda/U=0.3$, compared with the square of pairing gap $\Delta^{2}$ (blue dashed line). Inset: the corresponding pairing gap $\Delta$ as a function of $h$.}
\end{figure}

%\sectoc
The minima of pair-breaking energy are closely connected to the edge curves of dynamical structure factor in Fig. \ref{fig1} and \ref{fig2}. So we discuss them as a function of the transferred momentum along the high symmetry directions in BZ. In Fig. \ref{lambh}, the minimum energy of three pair-breaking excitations are shown.
First, the bottom magenta dotted line is the minimum energy of the $\{11\}$-type excitation. It has an obvious excitation gap between the zero energy and decreases as $h$ increasing in the BCS superfluid. However, in the topological superfluid, the excitation gap becomes very small in most transferred momentum, except for $[\pi,\pi]$ region, which leads to the appearance of a single-particle excitation peak after the collective excitation peak at ${\bf q}=[\pi,\pi]$. Second, the blue dashed line denotes the $\{12\}$ or $\{21\}$-type minimum energy needed to break a Cooper pair at a certain ${\bf q}$ from different branches of the spectrum. This excitation reflects the coupling between spin and orbital motion. Third, the red line corresponds to the minimum energy in the $\{22\}$-type excitations. Generally this signal is the weakest.
\section{IV. Zeeman field dependence of both $S_{peak}$ and $\Delta^2$}
\label{SI:scale}
Our results show that both $S_{peak}$ and $\Delta^{2}$ display an almost the same tendency for different $h$.
\end{document}